\documentclass[11pt]{amsart}
\usepackage{geometry}                
\usepackage{graphicx}
\usepackage{amssymb}
\usepackage{epstopdf}
\newcommand{\mc}{\mathcal}

\usepackage[colorlinks]{hyperref}
\hypersetup{colorlinks=true,
	linkcolor=red,
	anchorcolor=green,
	citecolor=blue,
	urlcolor=green
}

\title{Reconstruction of financial networks for robust estimation of systemic risk} 
\author{Iacopo Mastromatteo}
\address{Iacopo Mastromatteo, \it{International School for Advanced Studies, via Beirut 2/4, 34014, Trieste, Italy}}
\author{Elia Zarinelli}
\address{Elia Zarinelli, \it{LPTMS, CNRS and Universit\'e Paris-Sud, UMR8626, B\^at. 100, 91405 Orsay, France}}
\author{Matteo Marsili}
\address{Matteo Marsili, \it{The Abdus Salam International Center for Theoretical Physics, Strada Costiera 11, 34014 Trieste, Italy}}

\thanks{The authors wish to thank F. Caccioli, F. Krzakala, Y. Sun, L. Zdeborova for very useful discussions. I.M. acknowledges support from
GDRE 224 GREFI-MEFI CNRS-INdAM}

\begin{document}

\begin{abstract}
In this paper we estimate the propagation of liquidity shocks through interbank markets when the information about the underlying credit network is incomplete.
We show that techniques such as Maximum Entropy currently used to reconstruct credit networks  severely underestimate the risk of contagion by assuming a trivial
(fully connected) topology, a type of network structure which can be very different from the one empirically observed.
We propose an efficient message-passing algorithm to explore the space of possible network structures, and show that a correct estimation of the network degree
of connectedness leads to more reliable estimations for systemic risk. Such algorithm is also able to produce maximally fragile structures, providing a practical upper bound
for the risk of contagion when the actual network structure is unknown. We test our algorithm on ensembles of synthetic data encoding some features of real financial networks
(sparsity and heterogeneity), finding that more accurate estimations of risk can be achieved.
Finally we find that this algorithm can be used to control the amount of information regulators need to require from banks in order to sufficiently constrain the reconstruction of financial networks.
%
\end{abstract}

\maketitle

\section{Introduction}
The estimation of the robustness of a financial network to shocks and crashes is a topic of central importance to assess the stability of an
 economic system. Recent dramatic events evidenced
 the fragility of many economies, supporting the claim that ``the worldÕs financial system can collapse like a row of dominoes'' \cite{sch}. As a result,
governments and international organizations became increasingly concerned about systemic risk. 
The banking system is thought to be a fundamental channel in the propagation of shocks to the entire economy: the economic distress of an insolvent bank
can be transmitted to its creditors by interbank linkages, thus a shock can easily propagate to the whole network.
Unfortunately detailed data on banks bilateral exposures is not always available, and institutions are often left with the
problem of assessing the resilience of a system to financial shocks by exploiting an incomplete information set.
In this framework the reconstruction of bilateral exposures becomes a central issue for the estimation of risk, and requires the application of
sophisticated inference schemes to obtain reliable estimations. Among several methods, a commonly used tool for this task is the so called {\em entropy maximization method} \cite{bla,wells,dar,van}. The main limitation of this procedure is that it assumes a market structure which can be quite different from the actual one: it tends to spread the debt as evenly as possible, without assuming any heterogeneity in the structure for the network \cite{mis}.
Unfortunately these assumptions lead to an undervaluation of the extent of contagion, as the measure of the vulnerability to financial contagion depends crucially on the pattern of interbank linkages. Stress-tests used to quantitatively analyze this dependence confirm this results both for simulated and real data, as shown in figures  \ref{fig:plot_furfine}, \ref{fig:plot_furfinePL} and in Ref.\ \cite{mis}.  
\\
In this paper we will introduce a message-passing algorithm to overcome this limitation, and to sample efficiently the space of possible structures for the network.
This method can be used to propose plausible candidates for the real network structure, and to produce worst case scenarios for the spread of financial contagion.
We remark that despite the high cardinality of the set of possible network structures ($\sim 2^{N^2}$), we are able to generate plausible configurations in a time which scales quadratically in the number of unknown entries of the liability matrix.

In section \ref{secFramework} we introduce the main concepts and define the problem of network reconstruction, while in \ref{secDenseReconstruction} we present the Maximum 
Entropy (ME) algorithm, a commonly used procedure to infer credit networks from incomplete datasets. In section \ref{secSparseReconstruction} we show the idea which allows our
algorithm to explore the space of network structures and extend the validity of ME. Section \ref{secStressTest} describes the stress-test which we employ to analyze the robustness
of financial networks, and in section \ref{secSimData} we apply all these ideas to synthetic datasets. In section \ref{secThreshold} we discuss the reliability of the reconstruction algorithm as a function of the policy adopted by regulatory institutions. Finally in section \ref{secConclusions} we draw our last conclusions.

\section{Framework}
\label{secFramework}
Let us consider a set of $N$ banks $\mc B = \{b_0,\dots ,b_{N-1}\}$, in which each bank in $\mc B$ may borrow to or lend money from other banks in $\mc B$. This structure is encoded in the so-called liability matrix $L$, an $N \times N$ weighted, directed adjacency matrix describing the instantaneous state of a credit network. Each element $L_{ij}$ denotes the funds that bank $j \in \mc B$ borrowed from bank $i \in \mc B$ (regardless of the maturity of the debt). We fix the convention that $L_{ij} \ge 0$ $\forall (i,j) \in \mc B \times \mc B$, $L_{ii}=0$ $\forall i \in \mc B$. With this definition, the expression $L_i^{\rightarrow}  = \sum_j L_{ij} $ represents the total credit which the institution $i$ possesses against the system (also known as out-strength), while  ${L_j^{\leftarrow} } = \sum_i L_{ij} $ represents the total debt owed by the institution $j$ to the environment (in-strength).\footnote{Without loss of generality we consider a  closed economy ($\sum_i L_i^{\rightarrow}  = \sum_j {L_j^{\leftarrow} }$), by using bank $b_0$ as a placeholder to take into account flows of money external to the system.}
This matrix contains information about the instantaneous state of a credit network, and it is sufficient to estimate the risk of contagion in many cases of practical relevance. Indeed
 one is often unable to obtain from empirical data the complete expression for the matrix $L$. Data are typically
extracted by a bank balance sheets or by institutional databases \cite{austriaci}, and partial informations have to be coherently integrated into a list of plausible
liability matrices.  In the following discussion, we will suppose that three different types of informations about $L$ are available, as typically reported in the literature \cite{uppe}:
\begin{enumerate}
\item{All the debts larger than a certain threshold $\theta$ are known. This allows us to rescale all the elements of $L$ by $\theta$, so that we consider without loss of generality
liability matrices for which all the unknown elements are bound to be in the interval [0,1]. We assume to have at most order $N$ elements exceeding such threshold.}
\item{We assume a certain set of entries (which we take to be of order $N$) to be known. This corresponds to banks or bank sectors for which some particular position needs
 to be disclosed by law.}
\item{The total credit $L_i^{\rightarrow} $ and the total debit ${L_j^{\leftarrow} }$ of each bank are known. Acceptable candidates for liability matrices need to satisfy a set of $2N$ linear constraints,
whose rank is in general $\mc R \leq 2N-1$ (due to the closed economy condition).}
\end{enumerate}
We remark that we have defined a set of constraints of order $N$ elements, which is too small to single out a unique candidate for the true unknown liability matrix. The possible solutions
 compatible with the observations define a space $\Lambda$, whose members we denote with $\hat L $.
Let $U $ be the set of not directly known  (i.e. non-fixed by to constraints of type (1) and (2)) entries of the liabilities matrix. Then those entries of the liability matrix (whose number is $M= |U| $) are real numbers subject to domain constraints (they must be in $[0,1]$) and linear algebraic constraints (the sum on the rows and on the
columns must be respected). The ratio $M/\mc R \ge 1$ controls the degree of underdetermination of the network, and is typically much larger than one.

\section{Dense reconstruction}
\label{secDenseReconstruction}
A possible procedure to study the robustness of a financial network when the complete information about the liability matrix is not uniquely specified, is to pick from the set of
candidate matrices $\Lambda$ a representative matrix, and to test the stability uniquely for the network specified by such $\hat L$. In this case a criterion has to be chosen to select a
particular matrix out of the $\Lambda$ space, by doing some assumptions about the structure of the true $L_{ij}$. A choice which is commonly adopted  \cite{bla,wells,dar,van} is
based on the maximum entropy criteria, which assumes that banks spread their lending as evenly as possible.
%
The problem becomes in this case that of finding a vector $\vec L = \{ L_\alpha\}_{\alpha \in  U} $  (the unknown entries of the liability matrix) whose entries satisfy the algebraic and domain constraints and minimize the distance with the uniform vector $\vec Q =\{ Q_{\alpha} \}_{\alpha \in \mathcal U}$ (such that $\forall \alpha \quad Q_\alpha = 1$), where the distance is quantified by the Kullback-Leibler divergence
\[
D_{KL}(\vec L, \vec Q) = \sum_\alpha L_\alpha \log \frac{L_\alpha}{Q_\alpha} \; .
\]
The minimization of such function is a standard convex optimization problem, that can be solved efficiently in polynomial time. In financial literature this algorithm is known with the
name of Maximal Entropy (ME) reconstruction. We remark that by using this algorithm no entry is exactly put to zero unless it is forced by the algebraic constraints.\footnote{This algorithm is
not the only possible choice to extract a representative matrix out from the set $\Lambda$. Indeed existing algorithms share with the ME the property of returning
solutions located in the {\it interior} of $\Lambda$. On the other hand, when choosing a point at random in a compact set in very high dimension $d$, it is very likely that the point will be very close to the boundary (i.e. at a distance of order $1/d$). Hence, it is reasonable to expect that typical feasible liability matrices are located on or close to the boundaries of $\Lambda$. \label{foot_typical}}
\\

\section{Sparse reconstruction}
\label{secSparseReconstruction}
\begin{figure}[tbp] 
   \centering
   \includegraphics[width=3.5in]{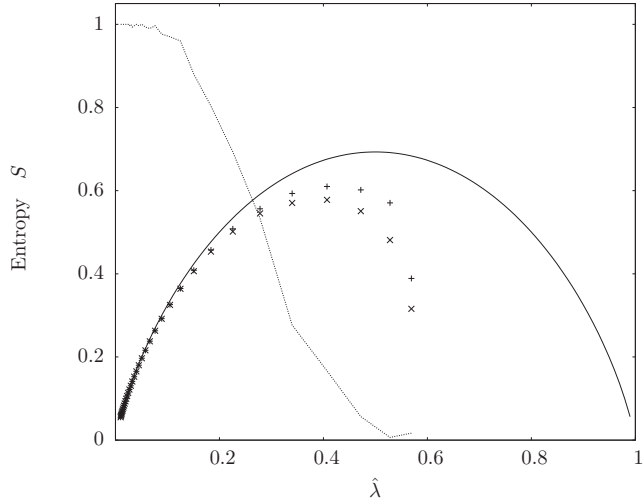} 
   \caption{Entropy $S$  of the space of compatible configurations $a_{i,j}$ at fixed sparsity $\hat \lambda$ with the energy $\mathcal{H}(a_{i,j})$ (+ sign) and
   true energy $\mathcal{H}_0(a_{i,j})$ ($\times$ sign) for the examples discussed in the text. $S$ is defined as the logarithm of the number of configurations $\{a_{i,j}\}$ with $\mathcal {H}=0$ (or $\mathcal {H}_0$), divided by the number $M$ of possibly non-zero entries $a_{i,j}$.
   The solid line plotted for comparison is the entropy of a system of independent links $a_{i,j}$ with the same density (i.e. number of non-zero links).
   The probability for a solution of $\mathcal{H}_0(a_{i,j})$ to be also a solution of $\mathcal{H}(a_{i,j})$ is also plotted on the same graph (dashed line).}
   \label{fig:Entropy Plot}
\end{figure}

ME might not be a particularly good description of reality since the number of counterparties of a bank is expected to be limited and much smaller than $N$,
while ME tends to produce
completely connected structures. In the case of real networks the degree of market concentration can be higher than suggested by ME. 
This systematically leads to an underestimation of risk, as a structure in which the debt is distributed homogeneously among the nodes is generally known to be able to absorb shocks more effectively than a system in which few nodes dominate the network \cite{mis}.
In order to be closer to reality and to estimate more accurately the risk contagion it is then necessary to reconstruct liability matrices whose degree of sparsity (i.e.\ the fraction of zero entries of $\hat L$) can be tuned, and eventually taken to be as big as possible. This corresponds to the choice of topologies for the interbank networks in which the number of links can be explicitly regulated by means of a control parameter. We present in this section an algorithm which, given the fraction $\hat \lambda$ of entries which are expected to be exactly zero, is able to reconstruct a sample of network structures compatible with this requirement, and to find a  $\lambda_{max}$ which bounds the maximum possible degree of sparsity.
We focus the discussion on the generic case in which topological properties of the original credit network such as the sparsity parameter $\lambda$ or the number of counter parties of each bank are not known, without imposing any specific type of null model. The purpose of the algorithm is to provide an efficient mean to explore the space $\Lambda$, and to illustrate how the result of the stress-testing procedures may vary according to the density of zeroes of the matrix $\hat L$ which is assumed. \\ \\
To be more specific, let us define the notion of \emph{support} of a liability matrix as follows: given an $N\times N$ weighted, directed adjacency matrix $L$, we define its support $a \in \{0,1\}^{N^2}$ as the $N \times N$ adjacency matrix such that
\[
a_{i,j} (L) = \Big\{ \begin{array}{ccc} 1 &\textrm{if}& L_{ij} > 0 \\ 0 && \textrm{otherwise}\end{array} \; .
\]
The \emph{sparsity} $\lambda$ associated with a specific network structure $a$ is defined as $\lambda\{ a_{i,j} \} = 1 - (\sum_{ij} a_{i,j})/N(N-1)$.  Finally, given a network structure  $\hat{ a}$ and a set of liability matrices $\Lambda$, we say that $\hat{ a}$ is \emph{compatible} with $\Lambda$ if there exists at least a matrix $\hat L \in \Lambda$ such that $a_{i,j} (\hat L) = \hat{a}_{i,j}$.
Now we consider a liability matrix $L$ which is partially unknown in the sense of section \ref{secFramework}, and address the following issues: (i) is it possible to fix a fraction $\hat \lambda$ of the unknown entries to zero without violating the domain and the algebraic constraints? More formally, this corresponds to ask whether it exists a matrix $\hat L \in \Lambda $ such that $\hat \lambda = \lambda(a(\hat L))$. More generally, (ii) how many supports $\hat{a}$ with fixed sparsity $\hat \lambda$ are compatible with $\Lambda$? 
The algorithm solves this problems by sampling from the space of all compatible supports $a(\Lambda)$ potential candidates whose degree of sparsity is constrained to be $\hat \lambda$, and by evaluating the volume of such support sub-space. As one can easily expect, there will be a range of $[ \lambda_{min}, \lambda_{max}]$ of fractions of fixed zeros compatible
with the constraints: trivially $\lambda_{min}=0$ corresponds to the dense network, which always admits a compatible solution, but we are able to find a non-trivial $\lambda_{max}$
which corresponds to the maximally sparse network of banks. A plot of the logarithm of the number of possible supports as a function of $\hat \lambda$ is given in figure \ref{fig:Entropy Plot} ($\times$ signs) for a
network as the ones described in section \ref{secSimData}.
Once a support is given, the liability matrix elements can easily be reconstructed via ME.\footnote{As shown in figure \ref{fig:plot_furfine} and \ref{fig:plot_furfinePL}, ME tends to underestimate the risk of contagion (see footnote \ref{foot_typical}) even in the case in which the true support $a(L)$ is known, thus suggesting that other reconstruction algorithms should be employed for the estimation of the non-zero entries of a partially known liability matrix. Indeed, it is clear from those same simulations that inferring the support corresponding to the original network is a significant first step towards a more correct estimation of the risk of contagion.}
\\
 
The algorithm that we use to sample the candidate network structures $\hat a$ employs a message-passing technique which is able to overcome the problem of explicitly inspecting the compatibility of each network. The main idea is that we want to associate to each adjacency matrix $a$ a sampling probability $P_0\{ a_{i,j} \}$, that is strictly zero for non-compatible supports and is otherwise finite. Sampling uniformly from the space of compatible supports would correspond to the choice that $P_0\{ a_{i,j} \} = 1 / |a(\Lambda)|$ iff $a_{i,j} \in a(\Lambda)$ and zero otherwise. Indeed, to fix the required degree of sparsity of the network $\hat \lambda$ one can consider the modified sampling probability
\[
P_0\{ a_{i,j} \} = \frac{1}{Z} \Big\{ \begin{array}{ccc} z^{\sum_{ij} a_{i,j}} &\textrm{if}& a \in a(\Lambda) \\ 0 && \textrm{otherwise}\end{array} \; ,
\]
where $Z$ is a normalization constant and the \emph{fugacity} $z$ controls the average degree of sparsity of the sampled network, and is fixed in order to recover $\hat \lambda = \sum_{a} P_0\{ a_{i,j} \}  \sum_{i\neq j} (1-a_{i,j})$. The variable $\log z$ is analogous to a chemical potential in physics, in the sense that it is used to select denser or sparser sub-graphs (i.e. tuning
the $\hat \lambda$ parameter). The probability distribution $P_0\{ a_{i,j} \} $ can also be seen as the $\beta \rightarrow \infty$ limit of
\begin{equation}
\label{exactProb}
P_0\{ a_{i,j} \} = \frac{1}{Z} e^{-\beta \, \mathcal{H}_0  \{ a_{i,j} \}} z^{\sum_{ij} a_{i,j}} \; ,
\end{equation}
where we introduce the formal cost function $\mathcal{H}_0 \{ a_{i,j} \}$ which vanishes for $a \in a(\Lambda)$ and is $1$ otherwise. Probability distributions of the form (\ref{exactProb}) are typically hard to compute explicitly due to the presence of the normalization constant $Z$, but their approximate marginals can be estimated efficiently (i.e.\ within a time scaling linearly in the number of unknown variables $M$) by means of the iterative algorithms such as the one described in the appendix. The solution that one obtains for the marginals
\begin{equation}
\label{marginals}
p_{0\,i,j} = \sum_{a} P_0\{ a_{i,j} \} \delta(a_{i,j} = 1)
\end{equation}
corresponds to the probability that the entry $a_{i,j}$ is equal to one in the ensemble of network structures which (i) are compatible with $\Lambda$ and (ii) have an average degree of sparsity $\hat \lambda$.
Being able to compute those marginals allow to sample efficiently the space of solution by employing procedures such as the \emph{decimation} one described in the appendix, in which at each step the most biased variable $(i,j)$ is fixed to $a_{i,j} = 1 $ with probability $p_{i,j}$, and a reduced problem in which such $a_{i,j}$ is held fixed is successively solved. Once all variables are fixed, an adjacency matrix $a_{i,j}$ is selected out of the space of solutions and can be used as a candidate network structure. \\ \\
Unfortunately, the energy function $\mathcal{H}_0$ is hard to manipulate, and we need to resort to an approximate energy function $\mathcal{H}$, whose structure is derived in the following paragraph.
Suppose that a liability matrix with unknown entries is given, together with of the vectors of total credit ($L_i^{\rightarrow} $) and the one of total liabilities (${L_i^{\leftarrow} }$).
Then without loss of generality one can assume the known entries to be equal to zero, as the values of the known entries can always be absorbed into a rescaled value of the $L_i^{\rightarrow} $ and
${L_i^{\leftarrow} }$, and the problem can be restricted just to the unknown entries of the matrix. Under this assumption we can define the set of banks $\mathbf B \subseteq \mathcal B$ which are linked to the unknown entries of the liability matrix.
Each node of $\mathbf B$ is a bank and the directed edges are the elements of $U$. For each node $i$ of $\mathbf B$ the sum of the incoming entries $L_i^{\rightarrow}  = \sum_j L_{ij}$ and of the outgoing entries  $L_j^{\leftarrow} = \sum_i L_{ij}$ is known.
Let $k_i^{\leftarrow}$ ($k_i^{\rightarrow}$) be the number of incoming (outgoing) links in the subset of edges where $L_{i,j}>0$. Since $L_{i,j}\le 1$, the number $k_i^{\leftarrow}$ ($k_i^{\rightarrow}$) of incoming (outgoing) links is at least the integer part of $L_i^{\leftarrow}$ ($L_i^{\leftarrow}$) plus one. Therefore, one can define a cost function\footnote{Here $\theta(x)=0$ for $x<0$ and $\theta(x)=1$ otherwise is the Heaviside step function.}
\begin{equation}
\label{eqApproxEnergy}
\mathcal{H}\{a_{i,j}\}=\sum_{i}\left[
\theta\left(L_i^{\rightarrow}-k_i^{\rightarrow}\right)+\theta\left(L_i^{\leftarrow}-k_i^{\leftarrow}\right)
\right]
\end{equation}
over the dynamical variables $a_{i,j}=0,1$ which identify the subset of edges, with
\[
k_i^{\rightarrow}=\sum_{j}a_{i,j}, \qquad k_i^{\leftarrow}=\sum_{j}a_{j,i} \; .
\]
Then we can construct the probability function
\begin{equation}
\label{approxProb}
P\{ a_{i,j} \} = \frac{1}{Z} e^{-\beta \, \mathcal{H}  \{ a_{i,j} \}} z^{\sum_{ij} a_{i,j}}
\end{equation}
which we employ to sample the space of candidate network structures. Notice that all sub-graphs $a_{i,j}$ with $\mathcal{H}=0$ are feasible candidates for the support of solutions $L_{i,j}>0$ to the problem. In general, the constraints are $2N$ linear equations and, as long as the number on non-zero elements $L_{i,j}$ is larger than $2N$ solutions exist, but it is not granted that they have $L_{i,j}\in [0,1]$ for al $i,j$.
In other words, all the compatible solutions have to satisfy the constraint $\mathcal{H}=0$, but the converse is not true (as shown in figure \ref{fig:Entropy Plot}), because some 
support $a_{i,j}$ may not admit a solution with $L_{i,j}\in [0,1]$ for al $i,j$. Equivalently, the cost function $\mathcal{H}_0 \{ a_{i,j} \}$  involves constraints that the approximate $\mathcal{H}$ is not able to capture.
\\
Message passing algorithms can be derived along the lines of Refs.\ \cite{pretti,zde} to solve efficiently the problem of sampling the space of solutions of (\ref{eqApproxEnergy}) as described in detail in appendix. In particular we propose a generalization the algorithm employed in Ref.\ \cite{pretti}, in which we consider hard constraints enforced by inequalities rather than equalities and add a fugacity parameter $z$ in order to control the density of links of the solutions.

\section{Furfine stress-test}
\label{secStressTest}
The aim of this section is to show that some measures of vulnerability of a banking system to financial contagion, also known with the name of stress-tests, are sensitive to the way in which the liability matrix is reconstructed. In particular the dense ME reconstruction typically underestimates the risk of contagion, while more realistic results are found if one employs a sparsification parameter $\hat \lambda$ controlling the density of links in a financial system.
\\

A widely used measure of vulnerability in financial literature is the stress-test introduced by Furfine \cite{fur}, which is a sequential algorithm to simulate contagion. Suppose that the liability matrix $L$ is given and let us define $C_z$ the initial capital of a bank $z$ in the system $\mc B$. The idea of the algorithm is simple: suppose that a bank $z$ of the ensemble $\mc B$ fails due to exogenous reasons. Then it is assumed that any bank $i \in \mc B$ loses a quantity of money equal to its exposure versus $z$ ($L_{iz}$) multiplied by an exogenously given parameter $\alpha \in [0,1]$ for loss-given-default. Then if the loss of the bank $i$ exceeds its capital $C_i$, bank $i$ fails. This procedure is then iterated until no more banks fail, and the total number of defaults is recorded.
\\
The procedure described above can be formally rephrased in the following steps: 
\\

\textbf{Step 0}: A bank $z \in B$ fails for external reasons. Let us define $D_0 =\{ z \}$, $S_0 = {\mc B} \backslash \{z\}$. For the banks $i \in S_0$ we set $C_i^0 =C_i$.
\\

\textbf{Step $t$}: The capital $C_i^{t-1}$ at step $t-1$  of banks $i \in S_{t-1}$ is updated according to 
\begin{equation}
C_i^{t} = C_i^{t-1} - \alpha \sum_{j \in D^{t-1}} L_{ij} \nonumber 
\end{equation} 
with $\alpha \in [0,1]$. A bank $i \in S_{t-1}$ fails at time $t$ if $C_i^t < 0$. Let us define $D_t$, the ensemble of all the banks $i \in S_{t-1}$ that failed at time $t$ and $S_t = S_{t-1} \backslash D_t$ the ensemble of banks survived at step $t$.
\\

\textbf{Step $t_{stop}$}: The algorithm stops at time $t_{stop}$ such that $D_{t_{stop}} = \varnothing$.  
\\

We remark that the capital $C_i$ of each bank is exogenously given, and in principle it is not linked to the liability matrix $L$. The same holds for $\alpha$, so that the result of
a stress-test is understood as a curve quantifying the number of defaults as a function of the $\alpha$ parameter.
Finally, the results of the stress-test depend on the first bank $z \in \mc B$ which defaults. Then one may choose either to consider the results of the stress-test dependent on the
$z$ which has been chosen or to average the outcome on all the banks in the system $\mc B$; we adopt this second type of measure, and consider the default of all the banks
to be equally likely.

\section{Application to synthetic data}
\label{secSimData}

\begin{figure}[tbp] 
   \centering
   \includegraphics[width=3in]{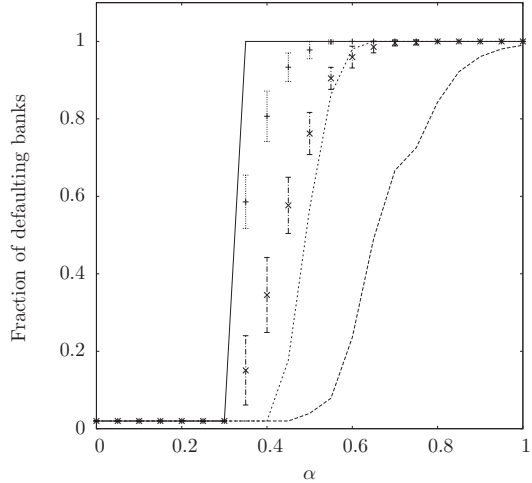} 
   \caption{Plot of mean fraction of failed banks vs loss-given-default parameter $\alpha$. The mean is done by averaging over the defaulting bank which starts the contagion. Results are obtained by considering: true liability matrix (solid line), reconstructed via ME algorithm liability matrix (thick dashed line) and the maximally sparse matrix ($+$ signs). Plots were obtained for a network of $N=50$ banks with entries uniform in [0,1], where the link probability was fixed to $0.7$ and the initial capital was set to $C_i= C = 0.3$. One can easily see that a  better estimation of the true risk of contagion is obtained if the reconstruction of the liability matrix is done by enforcing the correct sparsity of the network rather than with the ME algorithm: the results obtained by putting the \emph{correct} support (soft dashed line), corresponding to the original network structure $a(L_{ij})$, are also plotted, as well as the ones obtained by using a \emph{typical} support ($\times$ signs), corresponding to the choice of a random, compatible support $a_{i,j}$ whose degree of sparsity matches the one of the original network. Errors bars refer to the fluctuations of the default ratio associated with the choice of a specific support out of the ensemble the compatible ones at fixed degree of sparsity.}
   \label{fig:plot_furfine}
\end{figure}

In this section we will show how our algorithm of reconstruction of the liability matrix $L_{ij}$ (presented in section \ref{secSparseReconstruction}) gives more realistic stress-test results if compared with ME reconstruction algorithm (presented in section \ref{secDenseReconstruction}). 
\\
We choose to present the results obtained for specific ensembles of artificial matrices, whose structure should capture some of the relevant features of real credit networks\footnote{Our attempts to obtain data on real financial networks, such as those in Refs.\ \cite{mis,austriaci}, from central banks were unsuccessful. We focus on ensembles of homogeneous networks (i.e. non-scale free). This is appropriate since the unknown part of the financial network concerns small liabilities, and there is no a priory reason to assume a particularly skewed distribution of degrees for the unknown part of the financial newtork.}. The first case
that we analyze is the simplest possible network with a non-trivial topology, namely the one in which
every entrance of the liability matrix $L_{ij}$ with $i\neq j$ is set to zero with probability $\lambda$, and otherwise is a random number uniformly chosen in $[0,1]$. We set the banks initial capital $C_i$ to random numbers uniformly chosen in $[C_{min}, C_{max}]$. We impose the threshold $\theta = 1$, which means that all the entrance of the liability matrix are unknown (a worst-case scenario). We then reconstruct the liability matrix via ME algorithm and via our algorithm trying to fix the fraction $\hat \lambda$ of zeroes equal
to $\lambda$. Then we stress-test via the Furfine algorithm the three liability matrices: the true one, the one reconstructed via ME algorithm one and the reconstructed by means of
our message-passing algorithm, varying the loss-given-default $\alpha$ in $[0,1]$. The results of our simulations are shown in figure \ref{fig:plot_furfine}. We clearly show that the ME algorithm underestimates the risk of contagion, while more realistic results are obtained if the original degree of sparsity $\lambda$ is assumed. 

Notice that even when the degree of sparsity is correctly estimated, stress tests on the reconstructed matrix still underestimate systemic risk. This is because the weights $L_\alpha$ on the reconstructed sub-graph are assigned again using the ME algorithm. This by itself produces an assignment of weights which is much more uniform than a random assignment of $L_{ij}$ on the sub-graph, which satisfies the constraints (see footnote \ref{foot_typical}).
As a result, the propagation of risk is much reduced in the ME solution.

\begin{figure}[tbp] 
   \centering
   \includegraphics[width=3in]{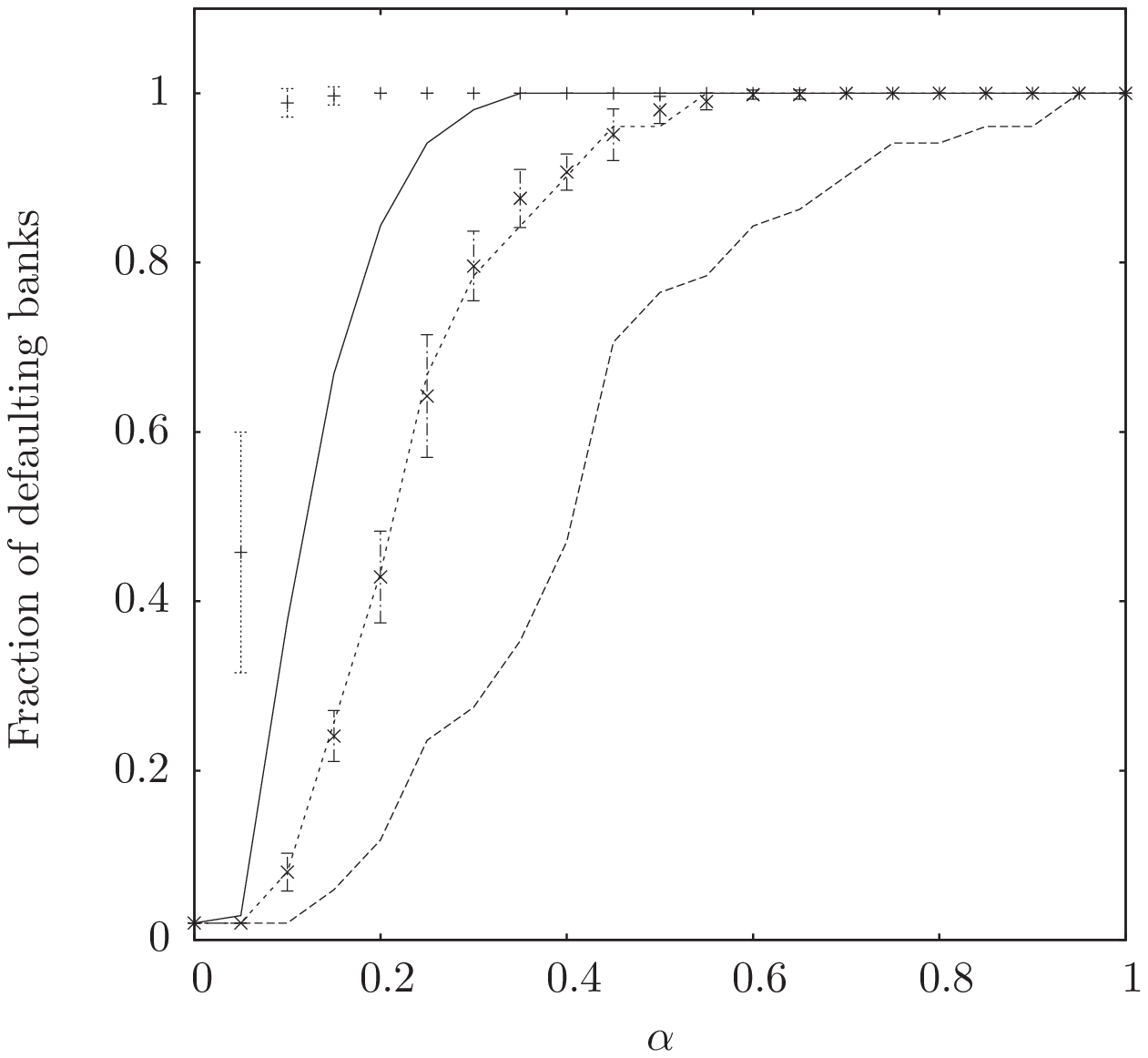} 
   \caption{A plot analogous to the one in figure \ref{fig:plot_furfine} for the case of power-law distributed entries of the liability matrix. This plots was obtained for a network of size $N=50$, where the link probability was fixed to $1/2$. The parameters for the distribution of the entries were set to $b=0.01$ and $\mu = 2$, while the capital of each bank was fixed to $C_i = C = 0.02$.}
   \label{fig:plot_furfinePL}
\end{figure}

The second ensemble that we consider is a simple extension of the first one, in which the only modification that we have introduced implements heterogeneity in the size
of the liabilities $L_{ij}$. In particular we consider matrix elements distributed according to
\[
p(L_{ij}) \sim (b+ L_{ij})^{-\mu-1} \; .
\]
Also in this case we can show (figure \ref{fig:plot_furfinePL}) that a more accurate estimation of the default probability is achieved by enforcing the sparsity parameter of the
reconstructed network to be the correct one. In this case the maximally sparse curve is less informative than in the uniform case. This is easily understood as due to
the fact that the typical element $L_{ij} \sim 10^{-2}$ is much smaller than the threshold $\theta = 1$, so that a number of zero entries substantially larger than the original one can be
fixed without violating the hard constraints.

In both cases, when the true sparsity of the network is unknown, focusing on the sparsest possible graph likely over-estimates systemic cascades, thereby providing a more conservative measure for systemic risk than the one obtained by employing ME alone.

\section{The role of the threshold}
\label{secThreshold}
\begin{figure}[tbp] 
   \centering
   \includegraphics[width=3in]{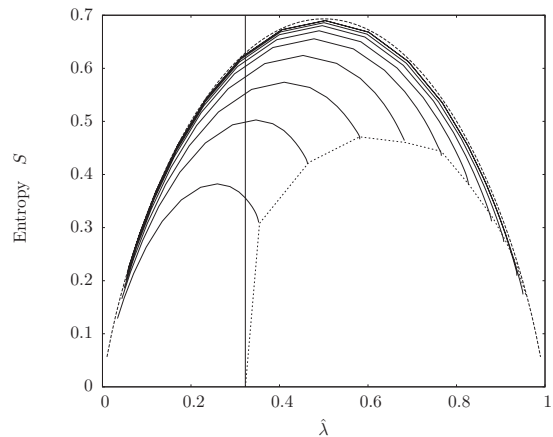} 
   \caption{We plot the entropy of the space of compatible distributions (i.e. of the solutions of $\mathcal H\{a_{i,j}\}$) as a function of the sparsity parameter $\hat \lambda$ by varying
   the threshold $\theta$ from 1 (top curve) to 0.01 (bottom curve). The dashed line signals the transition point where solutions cease to exist. We consider power-law distributed entries for the true network ($D=30$, $\lambda \approx 0.3$, $b=0.01$
   and $\mu = 2$). This shows how the volume of the space is reduced by a change of the threshold and how $\lambda_{max}$ gets closer to $\lambda$ by
   lowering $\theta$.}
   \label{fig:plot_EntrThresh}
\end{figure}

\begin{figure}[tbp] 
   \centering
   \includegraphics[width=3in]{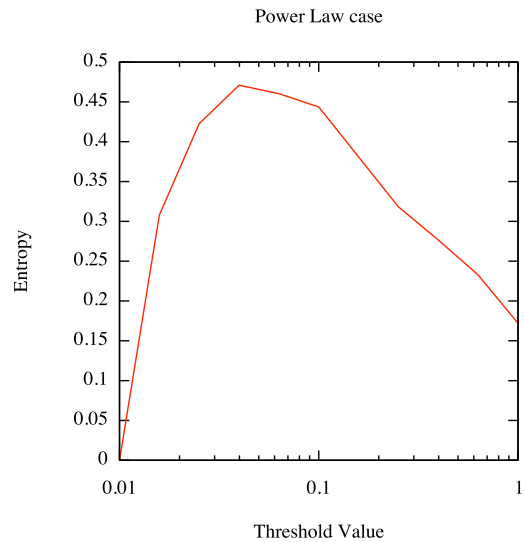} 
   \caption{The entropy of the space of solutions $\mathcal H\{a_{i,j}\}$ as a function of the threshold for the same network as the one depicted in figure \ref{fig:plot_EntrThresh}.}
   \label{fig:plot_CompatStruct}
\end{figure}

In the discussion above we disregarded the role of the threshold $\theta$ above which an exposure $L_{ij}$ has to be made publicly available to regulators by setting it equal to 1.
Indeed the problem of setting such threshold is a central problem to build a regulatory policy, hence the discussion of the reliability of the reconstruction algorithm varying $\theta$
while keeping fixed the true $L$ is in practice particularly relevant.
An appropriate way to address this issue is the following: given a network ensemble (such as the ones described in previous section) and a threshold $\theta$, how many network
structures are there with a compatible support? In particular, we remark that among all such compatible supports the maximally sparse one can be used to bound from above the
maximum amount of risk given a policy for the thresholding.
In particular for each value of $\theta$, we empirically find that $\lambda_{max}[\theta]$ enjoys the following properties:
\begin{enumerate}
\item{The maximum sparsity  $\lambda_{max}(\theta)$ is a decreasing function of $\theta$. In particular for $\theta \to 0$ one has $\lambda_{max}(\theta) \to \lambda$;}
\item{The entropy $S(\hat \lambda(\theta)) \to 0$ when the threshold goes to 0.}
\end{enumerate}
An example of this behavior for an ensemble of networks with power-law distributed weights is represented in figure \ref{fig:plot_EntrThresh}, while in \ref{fig:plot_CompatStruct}
we plot the entropy $S(\lambda_{max})$ structures as a function of $\theta$. 
Therefore the algorithm described in section \ref{secSparseReconstruction} provides quantitative measures for the uncertainty induced by the choice of a given threshold $\theta$ on network reconstruction. 
Ideally $\theta$ should be chosen so that maximally sparse structures are close to the true ones, and that the space of compatible structures is not too large (small entropy).

\section{Conclusions}
\label{secConclusions}
We have shown how it is possible to estimate the robustness of a financial network to exogenous crashes by using partial information. We confirm \cite{mis} that systemic risk measures depend crucially on the topological properties of the underlying network, and we show that the number of links in a credit network controls in a critical manner its
resilience: connected networks tend to absorb the response to external shocks more homogeneously than sparse ones.
We have also proposed an efficient message-passing algorithm for the reconstruction of the topology of partially unknown credit networks, in order to estimate with more accuracy
their robustness. Such algorithm allows (i) to sample the space of possible network structures, which is assumed to be trivial in Maximal Entropy algorithms
commonly employed for network reconstruction, and (ii) to produce typical credit networks, respecting the topological constraint on the total number of links.
Finally, we test our algorithms on ensembles of synthetic credit networks which incorporate some of the main features of real credit networks (sparsity and heterogeneity), and find
that the quality of the stress-test when only partial information is available critically depends on the assumptions which are done about the network topology.\footnote{
A similar problem of deriving network ensembles satisfying given constraints has been recently addressed in Ref.\ \cite{Garlaschelli}. However, Ref.\ \cite{Garlaschelli} focuses on a different problem, which is that of computing efficiently expected values of network properties in maximum entropy ensembles of networks with the same expected degree sequence of a given graph. Here we focus on the problem of deriving ensembles of partially observed networks. We mention, in passing, that ensemble properties, including local ones, can be very efficiently computed within our framework from the fixed point of the message passing equations for the marginals $\mu_{a\to b}$, in spite of the fact that our constraints are more complex, as they involve inequalities in the degrees.
} In particular, we find that
ME underestimates the risk of contagion if the sparsity of the real ensemble is big enough, while our algorithm provides less biased estimates.
We remark that a worst case analysis of the topology is possible using the proposed algorithm, as we are able to produce the maximally sparse (hence, maximally
fragile) possible structure for the network. Further developments of this work are indeed possible, in particular the identification and the reconstruction of other relevant topological features of credit networks would be relevant for a more accurate estimation of the contagion risk.


\begin{appendix}
\label{appendix}
\section{Message-passing algorithm}
We describe here the algorithm which we use to sample the solution space of the energy function
\begin{equation}
\mathcal{H}\{a_{i,j}\}=\sum_{i}\left[
\theta\left(L_i^{\rightarrow}-k_i^{\rightarrow}\right)+\theta\left(L_i^{\leftarrow}-k_i^{\leftarrow}\right)
\right]
\nonumber
\end{equation}
which we derived along the line of \cite{pretti}. Specifically, given as an input an incomplete liability matrix, whose information is encoded into a set of $N$ in-strength $L_i^\leftarrow$, $N$ out-strength $L_i^{\rightarrow}$ and a set of $U$ unknown entries of cardinality $M=|U|$), we provide an algorithm which for any positive value of the fugacity $z$ returns an adjacency matrices $a_{i,j}$ sampled according to the probability distribution (\ref{approxProb})
\[
P\{ a_{i,j} \} = \frac{1}{Z} e^{-\beta \, \mathcal{H}  \{ a_{i,j} \}} z^{\sum_{ij} a_{i,j}} \; .
\]
The procedure that we describe can further be separated in two main tasks: (i) given a probability distribution of the form (\ref{approxProb}), finding an efficient mean to calculate marginals $p_{i,j}$ defined analogously to (\ref{marginals}) and (ii) given a fast algorithm to calculate marginals, using them to find an adjacency matrix $a_{i,j}$ distributed according to $P\{ a_{i,j} \}$.
\subsection{Calculation of the marginals}
The structure of the problem admits a graphical representation as a factor graph, in which $|U|$ variable nodes are associated to the $a_{i,j}$ degrees of freedom, while the
constraints are represented as factor nodes. In particular, there are $2N$ function nodes, labeled $a\in\{i\rightarrow,\leftarrow i,~i=1,\ldots,N\}$ each with $k_a$ variable nodes attached.
Let the variables be denoted $x_{a,b}=x_{b,a}=0,1$ with $a,b$ and let $\partial a$ be the set of neighbors of node $a$. Let $M=\frac{1}{2}\sum_a |\partial a|$ be the total number of variables.
For each variable $x_{a,b}$ we define the \emph{message} $\mu_{a\to b}$ as the reduced marginal
\[
\mu_{a\to b}=  \sum_x P\{x_{a,b}|\not b\} \delta(x_{a,b} = 1) \; ,
\]
where $P\{x_{a,b}|\not b\}$ denotes the restriction of the probability measure (\ref{approxProb}) to a problem in which the function node $b$ is absent. Such messages need to fulfill self-consistent relations (BP equations) \cite{mont} which can be written in terms of the statistical weights\footnote{Since $k_a$ can be as large as $N$, the direct computation of $V_{S\to a}^m$ involved in principle $2^{k_a}$ terms, which may be very large. A faster way to compute it is to use the recursion relation
\[
V_{S\to a}^m=(1-\mu_{b\to a})V_{S\backslash b\to a}^m+ \mu_{b\to a}V_{S\backslash b\to a}^{m-1},~~~~~\forall b\in S \; .
\]
In practice this allows one to build $V_{S\to a}^m$ adding one at a time the nodes in $S$. This procedure involves of order $m^2\le k_a^2$ operations.}
\[
V_{S\to a}^m=\sum_{U\in S: |U|=m} \prod_{b\in U}\mu_{b\to a}\prod_{c\in S\backslash U}(1-\mu_{c\to a})
\]
and they read
\begin{eqnarray}
\label{BP1}
\mu_{a\to b} &=& \frac{\sum_{m=L_a-1}^{k_a-1}z^{m+1}V^m_{\partial a\backslash b\to a}}{\sum_{m=L_a-1}^{k_a-1}z^{m+1}V^m_{\partial a\backslash b\to a}+\sum_{m=L_a}^{k_a-1}z^{m}V^m_{\partial a\backslash b\to a}}\\
 & = & \frac{V^{L_a-1}_{\partial a\backslash b\to a}+z W_{\partial a\backslash b\to a}}{V^{L_a-1}_{\partial a\backslash b\to a}+(1+z) W_{\partial a\backslash b\to a}} \nonumber \\
 W_{\partial a\backslash b\to a} &=& \sum_{m=L_a}^{k_a-1}z^{m-L_a}V^m_{\partial a\backslash b\to a} \; . \label{BP2}
\end{eqnarray}
Here $z$ is the fugacity of links, and controls the average degree of sparsity $\hat \lambda$ of the supports in the solution space.
For $z\to 0$ we obtain the equation for the sparsest possible graph
\[
\mu_{a\to b} = \frac{V^{L_a-1}_{\partial a\backslash b\to a}}{V^{L_a-1}_{\partial a\backslash b\to a}+V^{L_a}_{\partial a\backslash b\to a}} \; ,
\]
whereas for $z\to\infty$ we recover the maximally connected graph $\mu_{a\to b}=1$ for all $a$ and $b\in\partial a$.

Once the fixed point of Eqs. (\ref{BP1},\ref{BP2}) is found by iteration, for a given $z$, one can compute the marginals
\[
p_{a,b} =\frac{\mu_{a\to b}\mu_{b\to a}}{\mu_{a\to b}\mu_{b\to a}+(1-\mu_{a\to b})(1-\mu_{b\to a})}
\]
that link $(a,b)$ is present, and the entropy
\[
S(z)=\sum_a \log \sum_{m=L_a}^{k_a}V_{\partial a\to a}^{m}-\frac{1}{2}\sum_a\sum_{b\in\partial a}\log\left[\mu_{a\to b}\mu_{b\to a}+(1-\mu_{a\to b})(1-\mu_{b\to a})\right]
\]
To plot the number of solutions (or of different supports) as a function of the sparsity parameter $\hat \lambda$, and the associated entropy $\Sigma(\hat \lambda)$
one should use the fact that:
\[
e^{MS(z)}=\int_0^1 \, d\hat \lambda \, e^{M \Sigma (\hat \lambda)+M (1- \hat \lambda) \log z}
\]
and hence perform the back-Legendre transform.
\subsection{Decimation}
We describe in the following a \emph{decimation} procedure to generate the configurations $a_{i,j}$ once that the problem of computing marginals $p_{i,j}$ is controlled. For simplicity we choose to present a simple version of the algorithm, while more detailed description of this procedure and a discussion of its efficient variants can be found in reference \cite{mont}. \\ \\
\textbf{Step 0}: Define the set $U^{(0)} = U$, and the in- and out-strengths $L_a^{(0)} = L_a$. The candidate network structure is defined as $a_{i,j}^{(0)} = 0$ if $(i,j) \in U$ and $a_{i,j}^{(0)} = a(L_{ij})$ otherwise. \\
\textbf{Step $t+1$}: Find the marginals $p_{i,j}^{(t)}$ corresponding of the probability distribution $P^{(t)}\{ a_{i,j} \}$ associated to the reduced problem defined by the incomplete matrix of unknown entries $U^{(t)}$ and in- and out-strengths $L_a^{(t)}$. Select the most biased variable $(i^\star,j^\star) = \textrm{argmin}_{(i,j) \in U^{(t)}} \min [ p_{i,j}^{(t)}, 1-p_{i,j}^{(t)}]$ and set:
\begin{eqnarray*}
a_{i^\star,j^\star}^{(t+1)}  &=& 1 \; \textrm{with\,prob.} \; p_{i^\star,j^\star}\\
U^{(t+1)} &=& U^{(t)} \, \backslash \, (i^\star,j^\star) \\
L^{\rightarrow\,(t+1)}_i &=& L^{\rightarrow \, (t)}_i - a_{i^\star,j^\star} \\
L^{\leftarrow\,(t+1)}_i &=& L^{\leftarrow \, (t)}_i - a_{i^\star,j^\star} \\
\end{eqnarray*}
\\
\textbf{Step $t_{stop}$}: The algorithm stops at time $t_{stop}$ such that $U^{(t_{stop})} = \varnothing$. The candidate support $a_{i,j} = a_{i,j}^{(t_{stop})}$ so-obtained is distributed according to the probability distribution (\ref{approxProb}). \\

\end{appendix}

\end{document}